\documentclass[12pt]{iopart}

\usepackage{graphicx}% Include figure files
\usepackage{dcolumn}% Align table columns on decimal point
\usepackage{caption}
\usepackage{subfigure}

\expandafter\let\csname equation*\endcsname\relax
\expandafter\let\csname endequation*\endcsname\relax 
\usepackage{amsmath}
\usepackage{amssymb}
\usepackage{hyperref}
\usepackage{url}
\usepackage{xspace}

\newcommand {\rootsNN}  	{\ensuremath{\sqrt{s_{_{NN}}}}}
\newcommand {\roots}    	{\ensuremath{\sqrt{s}}}
\newcommand{\GeVc}			{\ensuremath{{\,\text{Ge\hspace{-.08em}V\hspace{-0.16em}/\hspace{-0.08em}}c}}\xspace}

\newcommand{\TeV}			{\ensuremath{{\,\text{Te\hspace{-.08em}V}}}\xspace}
\newcommand{\ptt}			{\ensuremath{p_{\mathrm{T}}}\xspace}
\newcommand {\pttrg}        {\ensuremath{p_\mathrm{T}^{\mathrm{trig}}}}
\newcommand {\ptass}        {\ensuremath{p_\mathrm{T}^{\mathrm{assoc}}}}
\newcommand {\npart}        {\ensuremath{\mathrm{N}_\mathrm{part}}}
\providecommand{\PYNEW} {\textsc{pythia8}\xspace}

\begin{document}
\title{Correlations and fluctuations measured by the CMS experiment in pp and PbPb}
\author{Wei Li for the CMS collaboration\footnote{For the full list of
CMS authors and acknowledgments, see appendix ``Collaborations''.}}
\address{
\vspace{2mm}
{\scriptsize
Massachusetts Institute of Technology, 77 Mass Ave, Cambridge, MA 02139-4307, USA\\
}}
\ead{davidlw@mit.edu}
\begin{abstract}
Measurements of charged dihadron angular correlations are presented in proton-proton (pp) and Lead-Lead (PbPb) collisions, over a broad range of pseudorapidity and azimuthal angle, using the CMS detector at the LHC. In very high multiplicity pp events at center-of-mass energy of 7\TeV, a striking "ridge"-like structure emerges in the two-dimensional correlation function for particle pairs with intermediate \ptt\ of 1--3\GeVc, in the kinematic region $2.0<|\Delta\eta|<4.8$ and small $\Delta\phi$, which is similar to observations in heavy-ion collisions. Studies of this new effect as a function of particle transverse momentum are discussed. The long-range and short-range dihadron correlations are also studied in PbPb collision at a nucleon-nucleon center-of-mass energy of 2.76\TeV, as a function of transverse momentum and collision centrality. A Fourier analysis of the long-range dihadron correlations is presented and discussed in the context of CMS measurements of higher order flow coefficients. 
\end{abstract}

%\maketitle

Measurements of dihadron azimuthal correlations~\cite{Adler:2002tq,star:2009qa,Adare:2006nr,phenix:2008cqb,
Alver:2009id,Alver:2008gk} have provided a powerful
tool to study the properties of the strongly interacting 
medium created in ultrarelativistic nuclear collisions.
At RHIC, dihadron azimuthal correlation measurements extending to large
relative pseudorapidities resulted in the discovery of a ridge-shaped correlation
in central AuAu collisions between particles with small relative azimuthal
angles ($|\Delta\phi| \approx 0$), out to very large relative pseudorapidities
($|\Delta\eta|$)~\cite{star:2009qa,Alver:2009id}. Recently, a striking 
ridge structure has also been observed in very high
multiplicity proton-proton (pp) collisions at a center-of-mass energy of 7\TeV
at the LHC by the Compact Muon Solenoid (CMS)
Collaboration~\cite{Khachatryan:2010gv}, posing new challenges to the
understanding of these long-range correlations.

Dihadron correlations for charged particles have been measured 
extensively by the CMS experiment in pp collisions at \roots\ = 7\TeV 
and PbPb collisions at a nucleon-nucleon center-of-mass energy
(\rootsNN) of 2.76\TeV over a large phase space. The nearly 4$\pi$ solid-angle
acceptance of the CMS detector is ideally suited for studies of both short-
and long-range particle correlations. A detailed description
of the CMS detector can be found in Ref.~\cite{JINST}. The dihadron 
correlations analysis technique has been described in Ref.~\cite{ref:HIN-11-001-PAS}.

Long-range, near-side ($\Delta\phi \approx 0$) ridge-like 
azimuthal correlations for $2.0<|\Delta\eta|<4.8$ have 
recently been observed for the first time in high multiplicity pp 
collisions at \roots\ = 7~TeV~\cite{Khachatryan:2010gv}. The novel 
structure resembles similar features observed in relativistic heavy-ion experiments.
This feature is most evident in the intermediate 
transverse momentum range of both $1<\pttrg<3$~\GeVc and 
$1<\ptass<3$~\GeVc. 
Following up the first observation of the ridge correlation 
structure in high multiplicity pp collisions, 
new results are presented to study the detailed
event multiplicity, transverse momentum and pseudorapidity 
gap ($\Delta\eta$) dependence of the ridge effect 
using the full statistics data collected in 2010. 
With a dedicated high multiplicity tracking
high-level trigger (HLT) setup, the CMS experiment has a unique
capability in studying this effect.

\begin{figure}[thb]
  \begin{center}
    \subfigure{\includegraphics[width=0.45\linewidth]{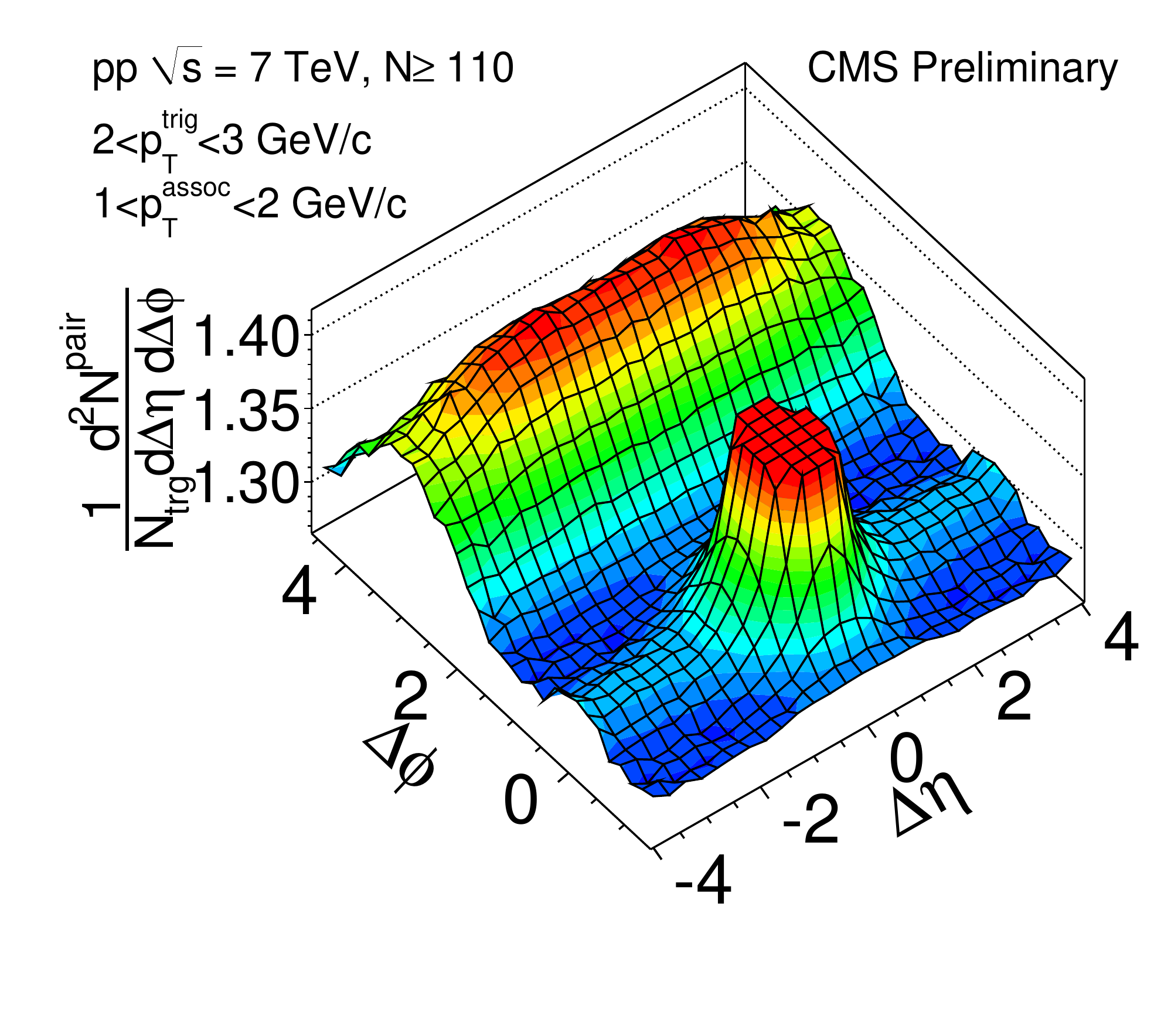}}
    \subfigure{\includegraphics[width=0.45\linewidth]{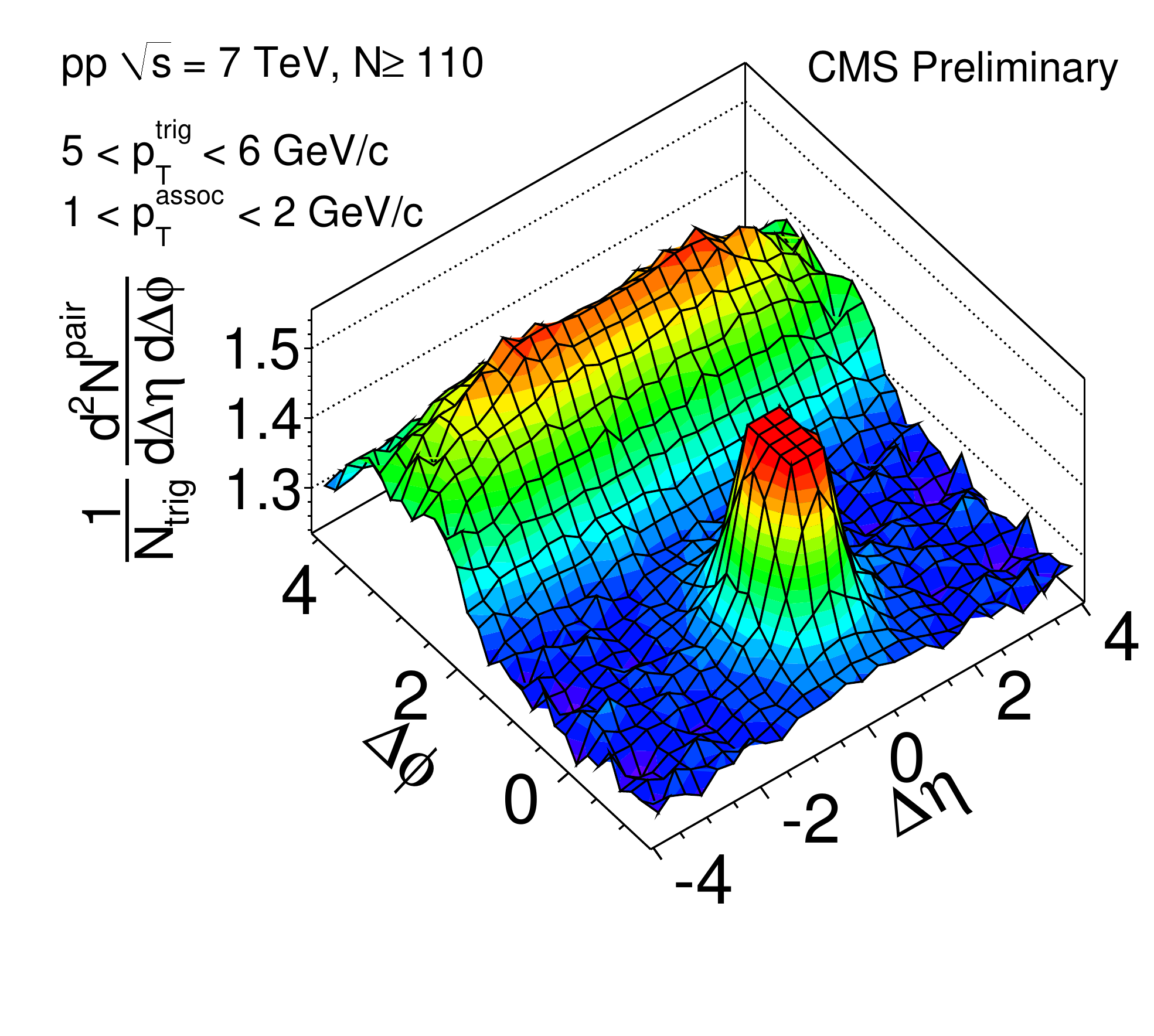}}
    \vspace{-0.8cm} 
    \caption{
         Two-dimensional (2-D) per-trigger-particle associated yield of charged hadrons
         as a function of $\Delta\eta$ and $\Delta\phi$
         with jet peak cutoff for better demonstration of the ridge from 
         high multiplicity ($N \geq 110$) pp collisions at \roots\ = 7~TeV, for 
         $2<\pttrg<3$ GeV/c and $1<\ptass<2$ GeV/c (left), and
         $5<\pttrg<6$ GeV/c and $1<\ptass<2$ GeV/c (right).
         }
    \label{fig:figure1}
  \end{center}
\end{figure}
\vspace{-0.8cm} 

The per-trigger-particle associated yield distribution of 
charged hadrons as a function of $\Delta\eta$ and $\Delta\phi$ 
in high multiplicity ($N \geq 110$) pp collisions at \roots\ = 7~TeV with trigger particles 
with $2<\pttrg<3~\GeVc$ and associated particles with $1<\ptass<2~\GeVc$ is shown in 
Fig.~\ref{fig:figure1} (left). The ridge-like structure is 
clearly visible at $\Delta\phi \approx 0$ extending to $|\Delta\eta|$ 
of at least 4 units as previously observed in Ref.~\cite{Khachatryan:2010gv}. 
However, at higher \pttrg\ of 5--6\GeVc as presented in Fig.~\ref{fig:figure1} (right), 
the ridge almost disappears.

\begin{figure}[thb]
  \begin{center}
    \subfigure{\includegraphics[width=\textwidth]{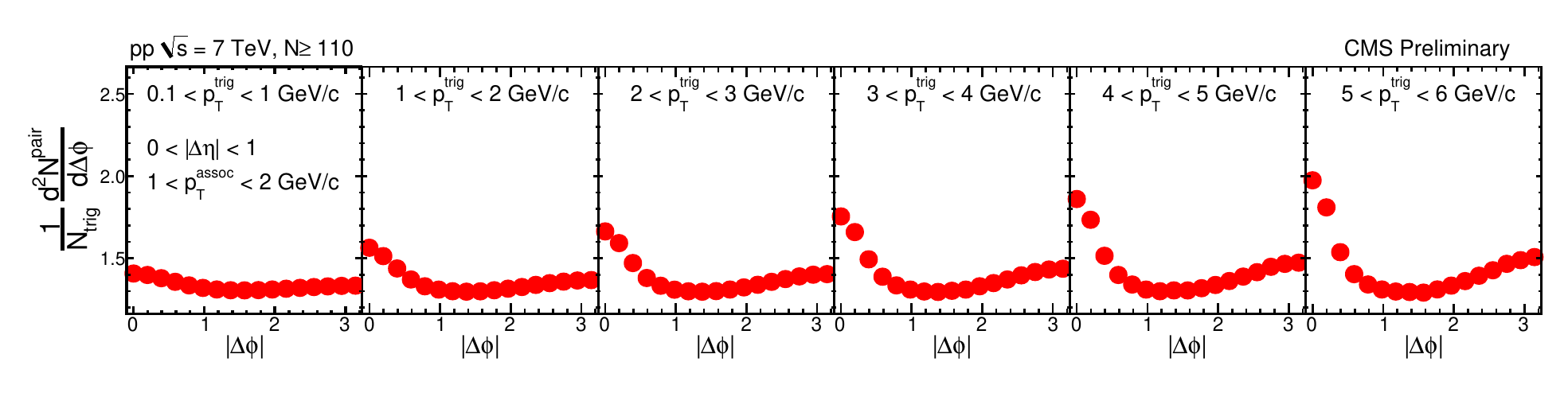}}
    \subfigure{\includegraphics[width=\textwidth]{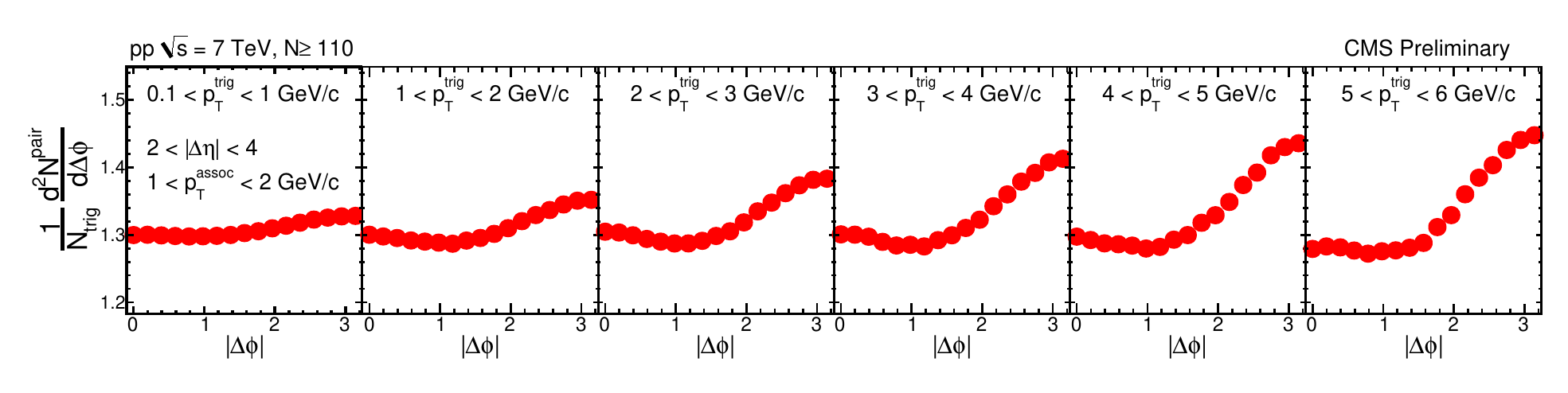}}
    \vspace{-0.8cm} 
    \caption{
        Short-range ($|\Delta\eta|<1$) and long-range ($2<|\Delta\eta|<4$) per-trigger-particle 
        associated yields of charged hadrons as a function 
        of $|\Delta\phi|$, from the high multiplicity ($N \geq 110$) 
        pp collisions at \roots\ =7~TeV, for six \pttrg\ and \ptass\ bins. 
        The error bars are statistical only
        and too small to be visible in most of the panels.
    }
    \label{fig:figure2}
  \end{center}
\end{figure}

In order to fully explore the detailed properties of both short-range jet-like 
correlations and long-range ridge-like structure, especially its dependence 
on event multiplicity and transverse momentum,
the associated yield distributions are obtained in eight bins ($2 \leq N < 35$, 
$35 \leq N < 45$, $45 \leq N < 60$,
$60 \leq N < 90$, $N \geq 90$,
$N \geq 110$, $N \geq 130$,
$N \geq 150$) of charged particle
multiplicity and six bins (0.1--1, 1--2, 2--3, 3--4, 4--5 and 5--6$~\GeVc$) 
of particle transverse momentum. The 1-D $\Delta\phi$ azimuthal correlation 
functions are calculated by integrating over the $0.0<|\Delta\eta|<1.0$ and 
$2.0<|\Delta\eta|<4.0$ region, defined as the jet region and ridge region, respectively. 
Fig.~\ref{fig:figure2} shows the results 
in the jet and ridge region respectively, for $N \geq 110$
in six bins of \pttrg\ at $1<\ptass<2$\GeVc.

\begin{figure}[thb]
  \begin{center}
    \subfigure{\includegraphics[width=0.45\linewidth]{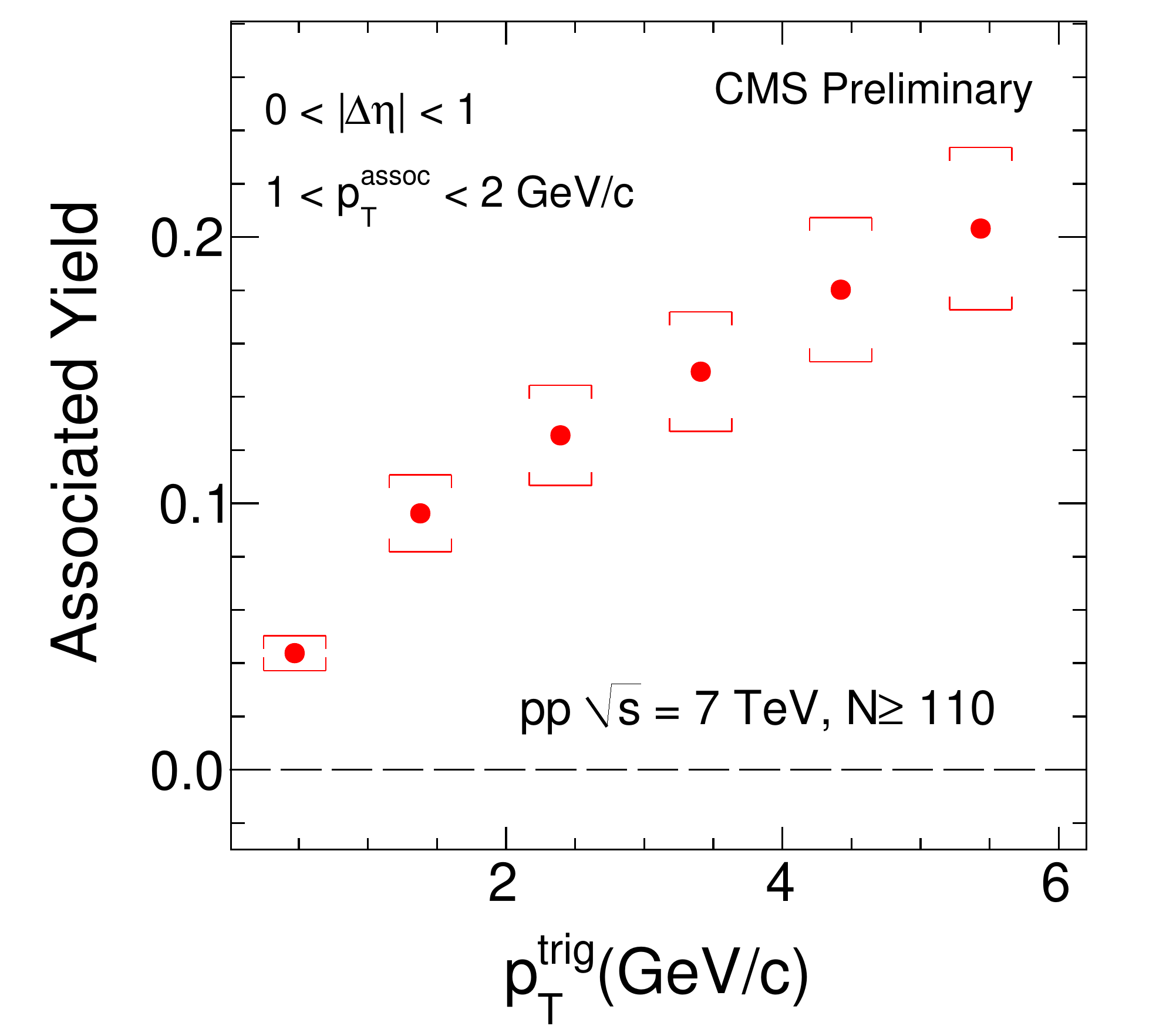} }
    \subfigure{\includegraphics[width=0.45\linewidth]{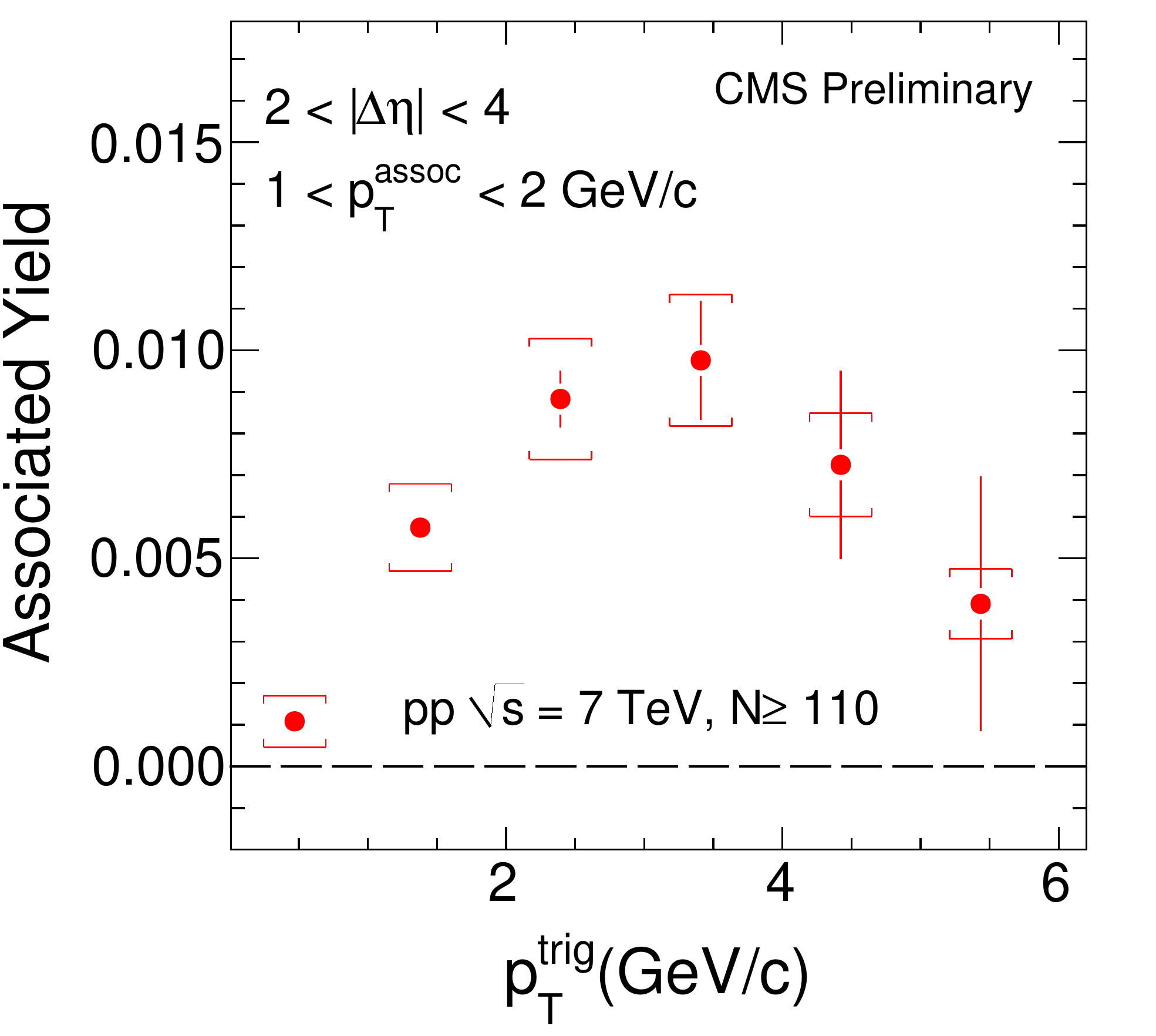} }
    \vspace{-0.5cm}        
    \caption{
        Integrated near-side
        associated yields for the short-range jet region ($0<|\Delta\eta|<1$) and
        the long-range ridge region ($2<|\Delta\eta|<4$),
        with $1<\ptass<2~\GeVc$,
        above the minimum level found by the ZYAM procedure, as a function 
        of \pttrg\ for $N \geq 110$ of pp collisions at \roots\ = 7~TeV. 
        The statistical uncertainties are shown as bars, while the brackets 
        denote the systematic uncertainties.
    }
    \label{fig:figure3}
  \end{center}
\end{figure}

In the next step, the near-side (small $\Delta\phi$ region) integrated associated 
yield is calculated for both jet and ridge regions relative to 
the minimum of the distribution. The details of this so-call "ZYAM" procedure can be 
found in Ref.~\cite{ref:HIN-11-001-PAS}. Figure~\ref{fig:figure3}
shows the integrated near-side associated yield for both the jet and ridge region
correlations with $1<\ptass<2~\GeVc$ (the \ptt\ range where the ridge effect 
appears to be strongest) as a function of \pttrg\ for $N \geq 110$.
The jet yield increases with \pttrg\ as expected due to 
the increasing contributions from high \ptt jets. The ridge yield first increases  
with \pttrg, reaches a maximum around \pttrg\ $\sim$ 2-3~\GeVc and 
drops at higher \pttrg.

\begin{figure}[thb]
  \begin{center}
    \subfigure{\includegraphics[width=0.45\linewidth]{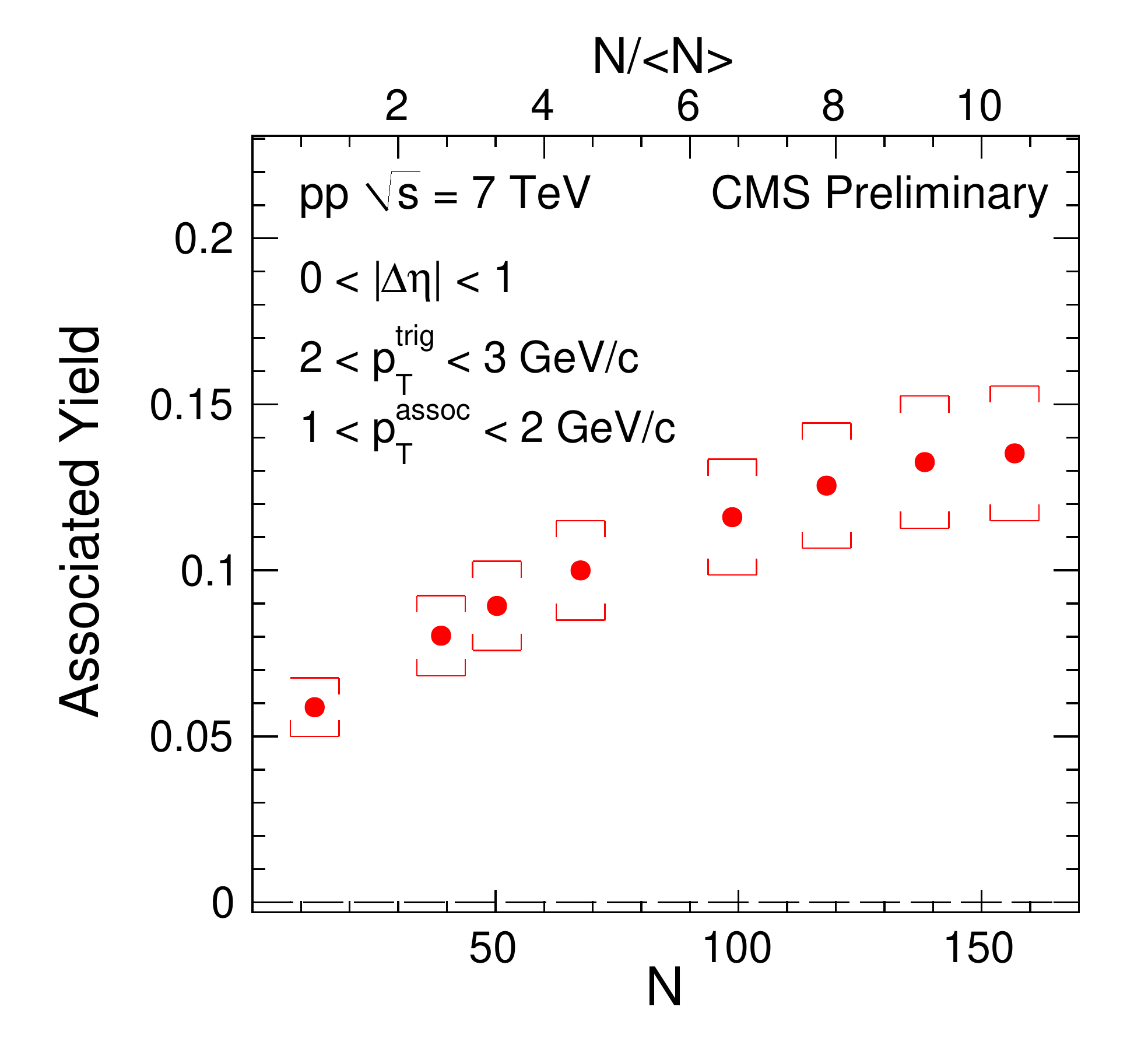} }
    \subfigure{\includegraphics[width=0.45\linewidth]{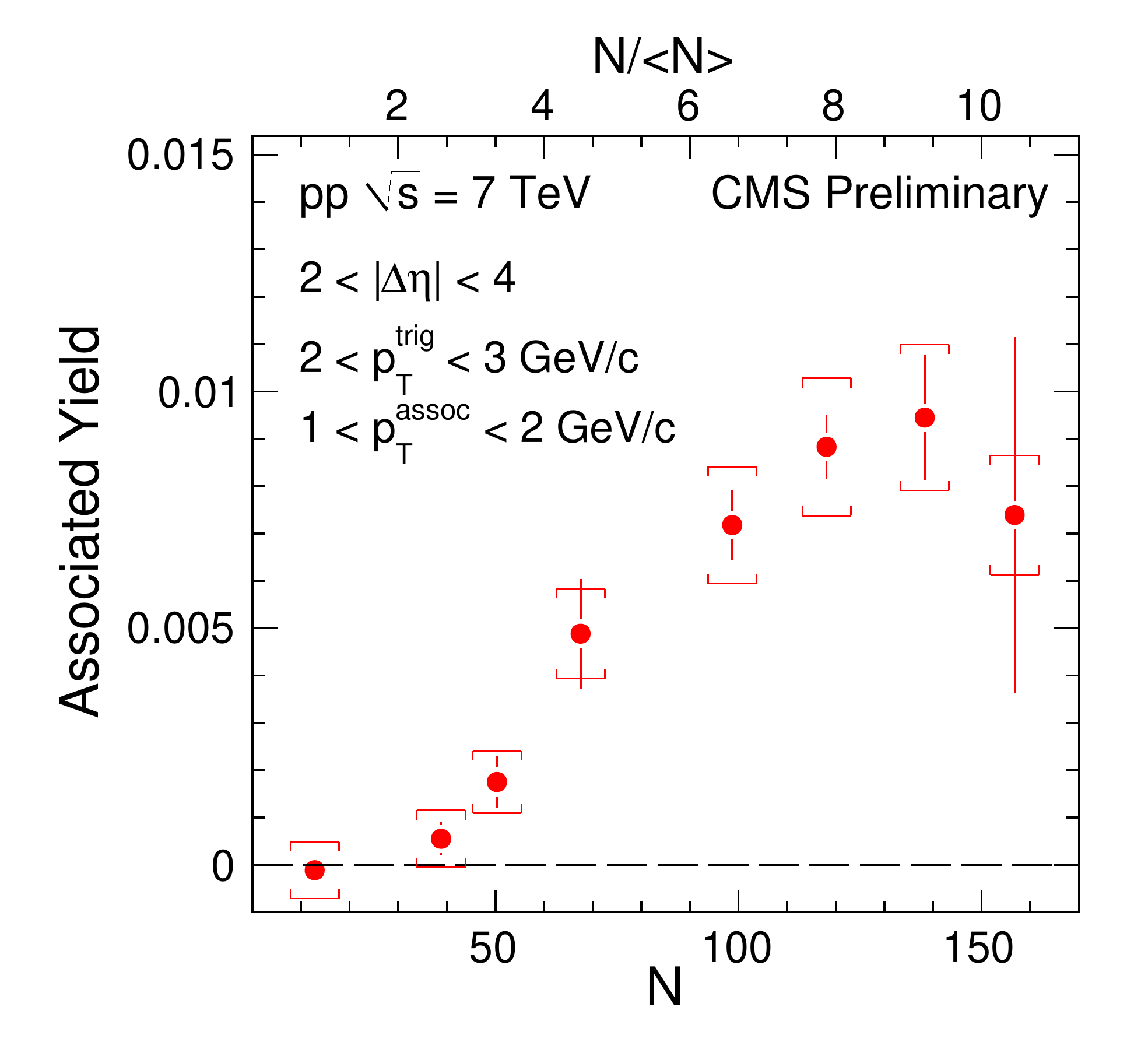} }
    \vspace{-0.5cm}        
    \caption{
        Integrated near-side
        associated yields for the short-range jet region ($0<|\Delta\eta|<1$) and 
        the long-range ridge region ($2<|\Delta\eta|<4$),
        with $2<\pttrg<3~\GeVc$ and $1 <\ptass< 2~\GeVc$,
        above the minimum level found by the ZYAM procedure, as a function 
        of event multiplicity from pp collisions at \roots\ = 7~TeV. 
        The statistical uncertainties are shown as bars, while the brackets 
        denote the systematic uncertainties.
    }
    \label{fig:figure4}
  \end{center}
\end{figure}

The multiplicity dependence of the near-side associated yield in the jet and ridge region
is illustrated in Fig.~\ref{fig:figure4} for one 
transverse momentum bin of $2<\pttrg<3~\GeVc$ and 
$1<\ptass<2~\GeVc$, the \ptt\ bin where the ridge effect appears to be strongest.
The ridge effect gradually turns on with event multiplicity around $N \sim 50-60$ (about 
four times of the average multiplicity in minimum bias events) and smoothly increases toward
high multiplicity region.

\begin{figure}[thb]
  \begin{center}
    \includegraphics[width=0.9\textwidth]{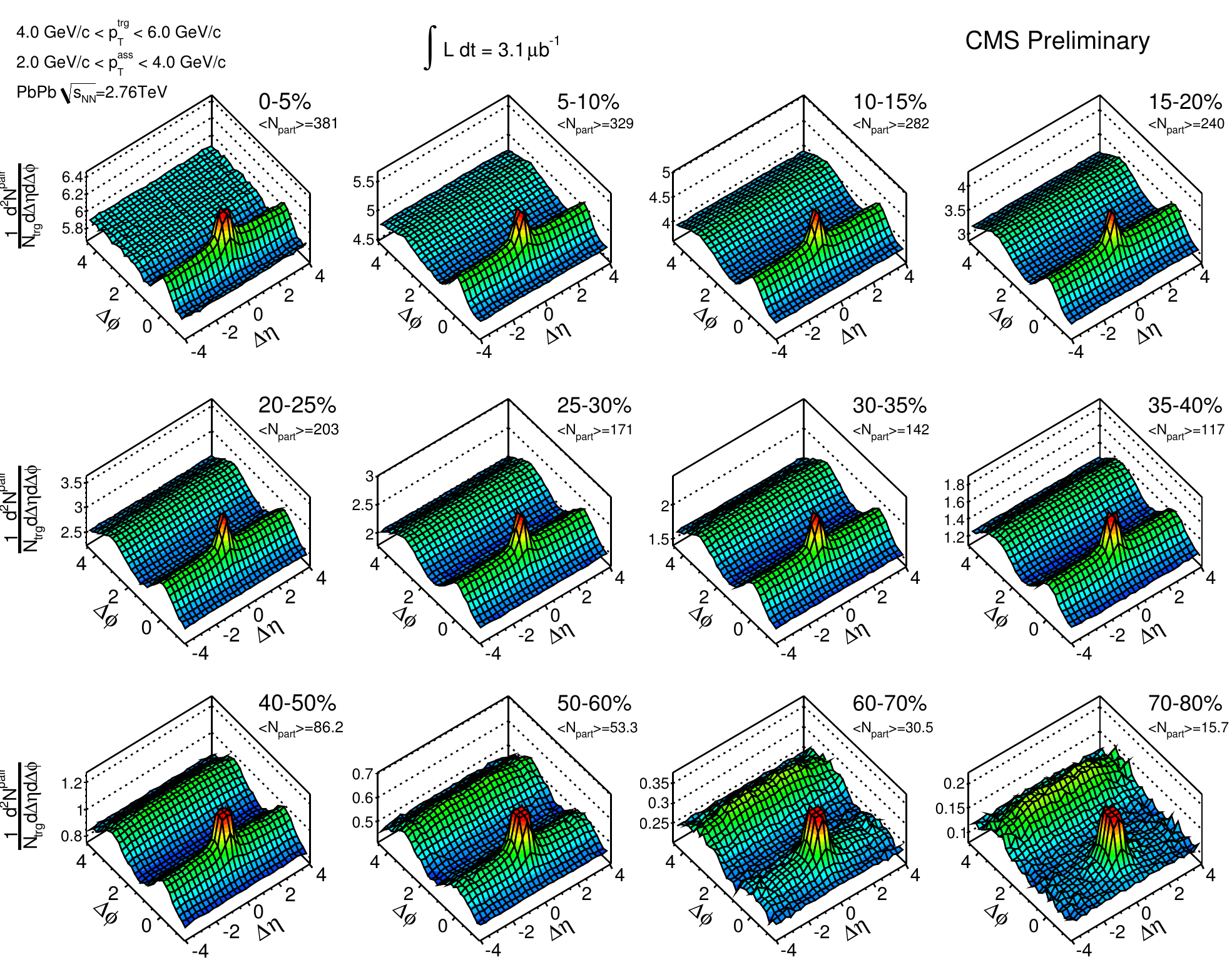}
    \vspace{-0.4cm}
    \caption{
         Two-dimensional (2-D) per-trigger-particle associated yield of charged hadrons
         as a function of $\Delta\eta$ and $\Delta\phi$ for 
         $4<\pttrg<6$~\GeVc and $2<\ptass<4$~\GeVc in 
         12 centrality classes of PbPb collisions at \rootsNN\ = 2.76~TeV. 
         The centrality labeling is such that 0-5\% is 
         the most central five percent of PbPb collisions.}
    \label{fig:figure5}
  \end{center}
\end{figure}

Moving onto PbPb collisions, dihadron correlations have been systematically studied as
a function of centrality and particle transverse momentum. An example of the 
2-D correlations for trigger particles with $4<\pttrg<6~\GeVc$ and associated particles 
with $2<\ptass<4~\GeVc$ is shown in Fig.~\ref{fig:figure5},
starting from 0-5\% corresponding to the most central collisions in the upper 
left-hand corner, to the 70-80\% bin corresponding to the most peripheral ones 
in the lower right-hand corner. In the 0--5\% most central PbPb collisions, 
a clear and significant ridge-like structure is observed at $\Delta\phi \approx 0$, which extends 
all the way to the limit of the measurement of $|\Delta\eta| = 4$. 
For more peripheral collisions, a $\cos(2\Delta\phi)$ component becomes 
prominent, which is attributed to the elliptic flow effect.

\begin{figure}[thb]
  \begin{center}
    \subfigure{\includegraphics[width=\textwidth]{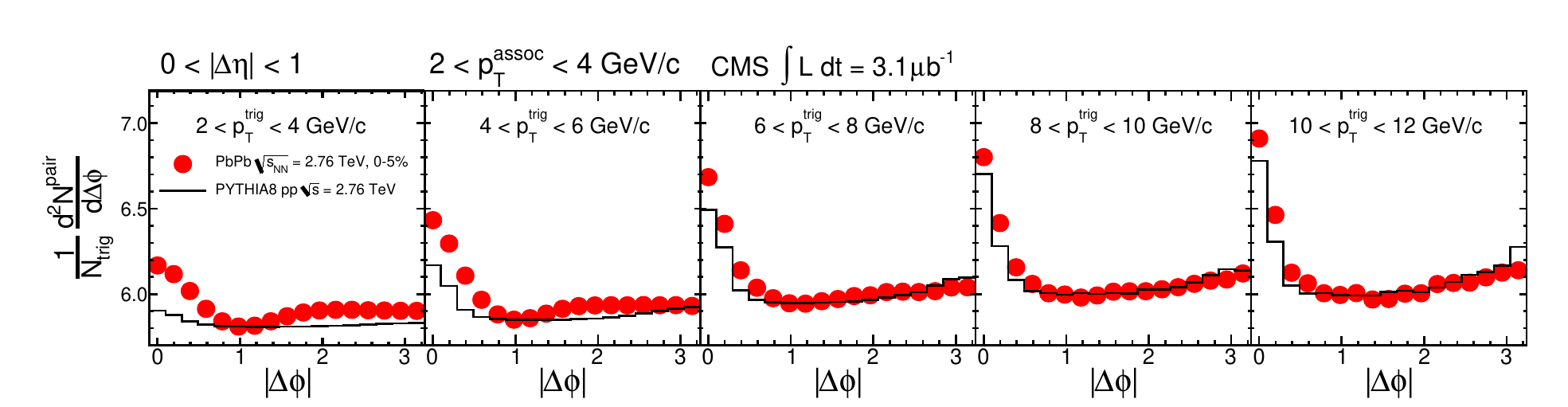}}
    \subfigure{\includegraphics[width=\textwidth]{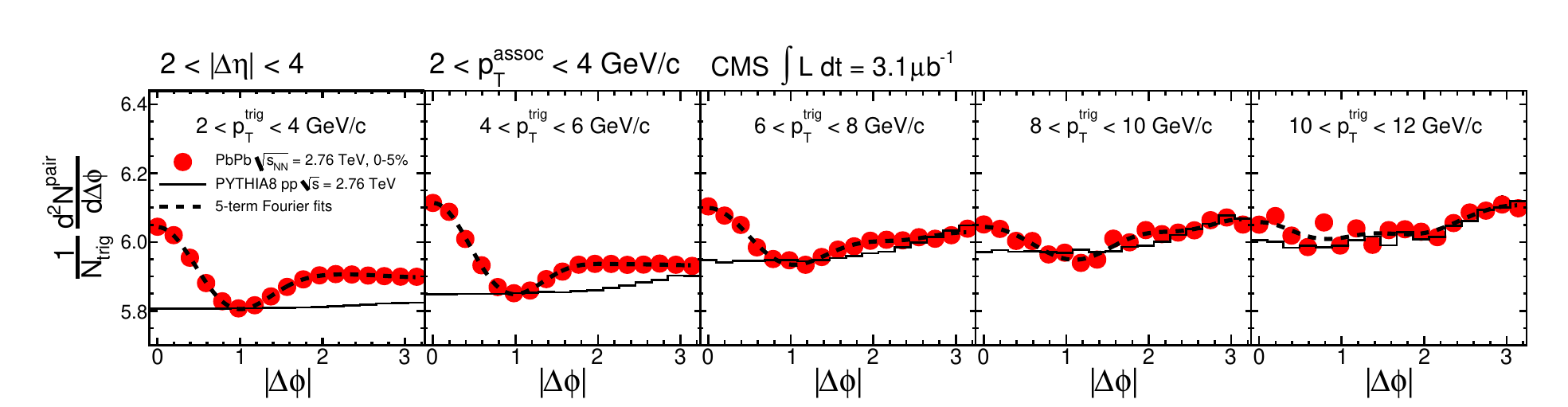}}
    \vspace{-0.5cm}        
    \caption{Short-range ($|\Delta\eta| < 1$) and long-range ($2 < |\Delta\eta| < 4$)
        per-trigger-particle associated yields of charged hadrons as a function 
        of $|\Delta\phi|$ from the 0--5\% most central PbPb collisions at \rootsNN\ = 2.76\TeV, 
        requiring $2 < \ptass < 4$\GeVc, for five different intervals of \pttrg.
        The \PYNEW pp MC results (solid histograms) are also shown, shifted up 
        by a constant value to match the PbPb data at $\Delta\phi=1$ 
        for ease of comparison. The error bars are statistical only
        and are too small to be visible in most of the panels. The systematic uncertainty
        of 7.6\% for all data points is not shown in the plots.}
    \label{fig:figure6}
  \end{center}
\end{figure}

The results of the 1-D $\Delta\phi$ correlations 
for the 0--5\% most central PbPb collisions 
are shown in Fig.~\ref{fig:figure6}.
The associated yield per trigger particle in the range of
$2 < \ptass < 4$\GeVc is extracted for
five different \pttrg\ intervals (2--4, 4--6, 6--8, 8--10, and 10--$12\GeVc$), and for both
jet and ridge region.
A comparison to \PYNEW pp MC events at $\sqrt{s}=2.76$\TeV 
is also shown, with a constant added to match the PbPb
results at $\Delta\phi=1$ in order to facilitate the comparison. 
In PbPb, the height of the ridge structure decreases as 
\pttrg\ increases and has 
largely vanished for $\pttrg \approx 10$--$12\GeVc$, a feature
that is qualitatively similar to what is observed in 
high multiplicity pp collisions, as was presented earlier.
The diminishing height of the ridge with increasing \pttrg\ 
was not evident from previous measurements in AuAu 
collisions. 

\begin{figure}[thb]
  \begin{center}
    \includegraphics[width=\linewidth]{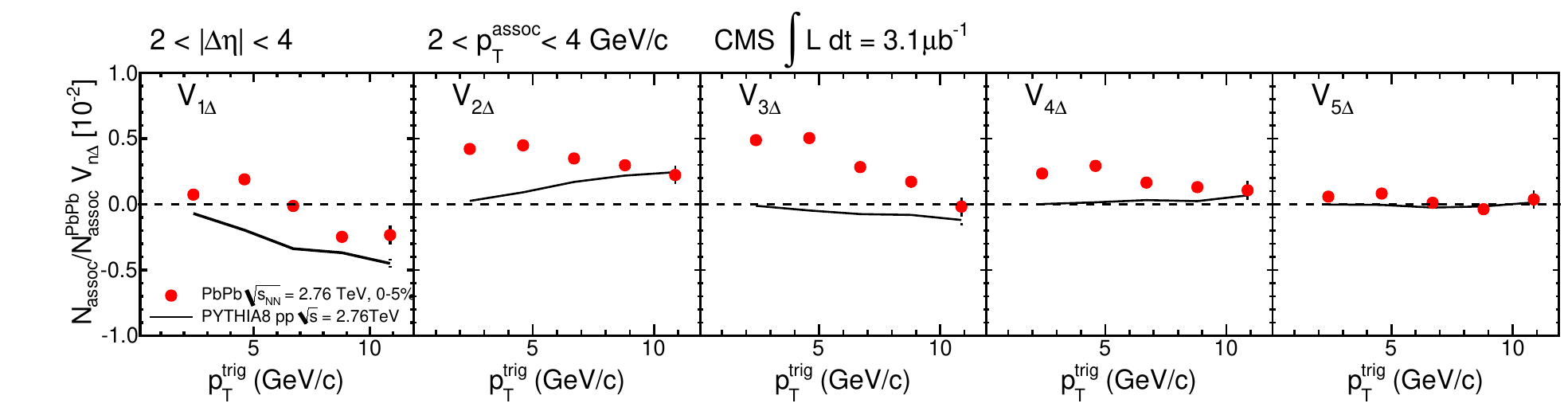}
    \vspace{-0.5cm}        
    \caption{Fourier coefficients, $V_{1\Delta}$, $V_{2\Delta}$, $V_{3\Delta}$, 
    $V_{4\Delta}$, and $V_{5\Delta}$, extracted
    as functions of \pttrg\ for $2 < \ptass < 4$\GeVc for the 0--5\% most central PbPb 
    collisions at \rootsNN\ = 2.76\TeV. The error bars represent statistical uncertainties only.
    The solid lines show the predictions from the \PYNEW 
    simulation of pp collisions at \roots\ = 2.76\TeV.}
    \label{fig:figure7}
  \end{center}
\end{figure}

Motivated by the idea of understanding the 
long-range ridge effect in the context of higher-order hydrodynamic flow 
induced by the initial geometric fluctuations~\cite{Alver:2010gr,Alver:2010dn,Schenke:2010rr,
Teaney:2010vd}, an approach of quantifying 
the observed long-range correlations using a Fourier decomposition technique is 
investigated. The 1-D $\Delta\phi$-projected distribution
is decomposed into Fourier series using the following expression:

\vspace{-0.2cm}
\begin{equation}
\label{fourier}
\frac{1}{N_{\rm trig}}\frac{dN^{\rm pair}}{d\Delta\phi} = \frac{N_{\rm assoc}}{2\pi} \lbrace 1+\sum\limits_{n=1}^{\infty} 2V_{n\Delta} \cos (n\Delta\phi)\rbrace,
\end{equation}
\vspace{-0.2cm}

\noindent 
where $N_{\rm assoc}$ represents the total number of dihadron pairs per trigger 
particle for a given $|\Delta\eta|$ range and (\pttrg, \ptass) bin. The 1-D $\Delta\phi$ 
projections are fitted by the first 
five terms in the Fourier series, the resulting fits being shown as the dashed 
lines in Fig.~\ref{fig:figure6}. The data are well described 
by this fit. The extracted first five Fourier coefficients from the fit are presented 
in Figure~\ref{fig:figure7} as functions of \pttrg\ for $2 < \ptass < 4$\GeVc
for the 0--5\% most central PbPb collisions and for the \PYNEW pp MC simulation. The 
\PYNEW results are scaled by the ratio of $N_{\rm assoc}^{\rm \PYNEW}/N_{\rm assoc}^{\rm PbPb}$
in order to remove the trivial dilution of the correlation strength with increasing multiplicity.

If the observed correlation was purely driven by the single-particle azimuthal
anisotropy arising from the hydrodynamic expansion of the medium~\cite{Voloshin:1994mz}, 
the extracted $V_{n\Delta}$ components would factorize into the flow coefficients 
$v_{n}$ (i.e., $v_{2}$ for anisotropic elliptic flow):

\begin{equation}
\label{eq:factorization}
V_{\rm n\Delta}(\pttrg, \ptass) = v^{\rm f}_{\rm n}(\pttrg) \times v^{\rm f}_{\rm n}(\ptass),
\end{equation}

\noindent where $v^{\rm f}_{\rm n}(\pttrg)$ and $v^{\rm f}_{\rm n}(\ptass)$ 
are the flow coefficients for the trigger and associated particles~\cite{Alver:2010gr}. 

\begin{figure}[thb]
  \begin{center}
    \includegraphics[width=0.6\linewidth]{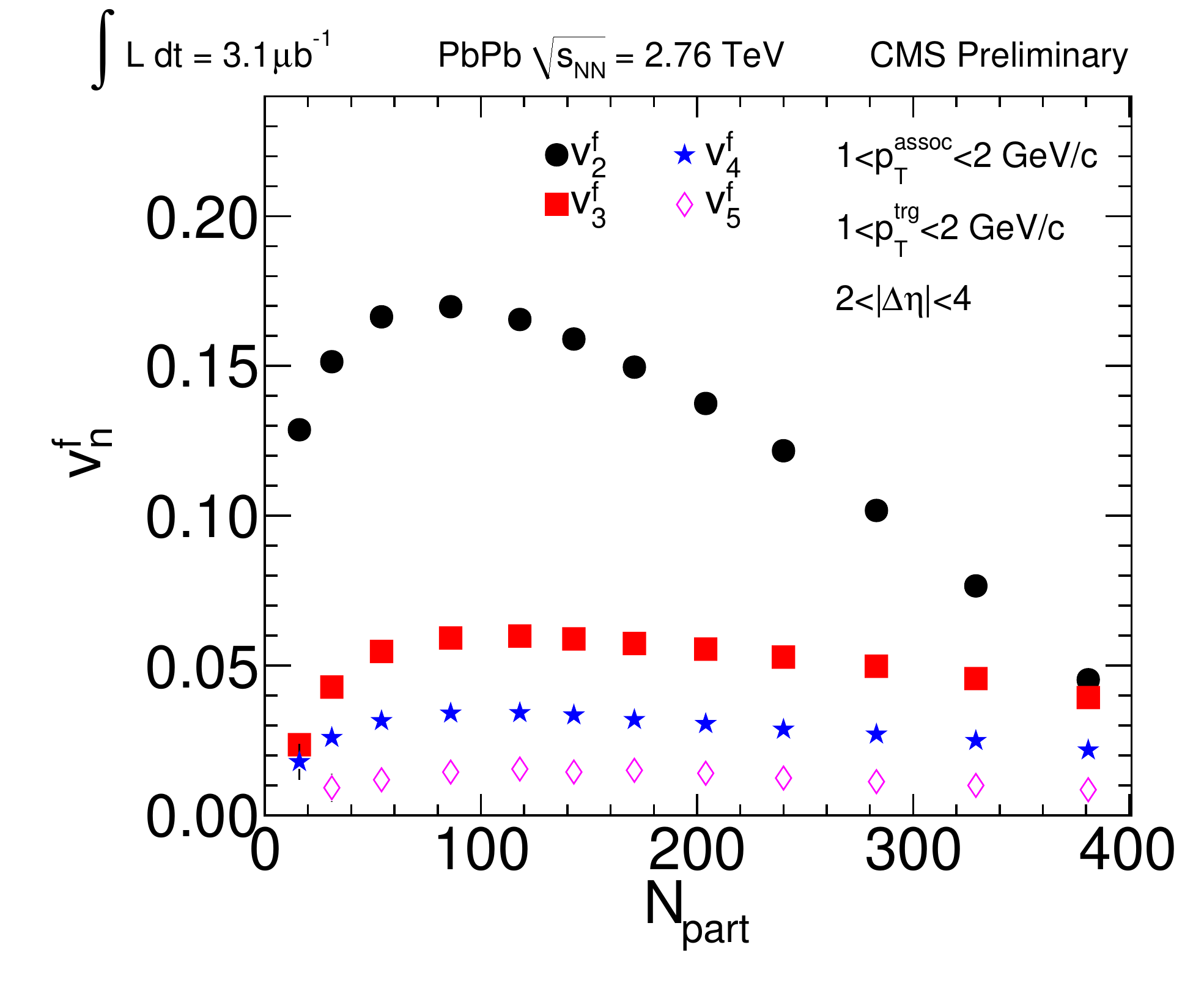}
    \vspace{-0.5cm}        
    \caption{
         The flow harmonics $v^{\rm f}_{2}$, $v^{\rm f}_{3}$, $v^{\rm f}_{4}$ and 
         $v^{\rm f}_{5}$ extracted from long-range ($2<|\Delta\eta|<4$) 
         azimuthal dihadron correlations for $1<\pttrg<2$~\GeVc and $1<\ptass<2$~\GeVc
         as a function of the number of participating nucleons (\npart) in PbPb collisions 
         at \rootsNN\ = 2.76\TeV.}
    \label{fig:figure8}
  \end{center}
\end{figure}
\vspace{-0.4cm}

Following this assumption in Eq.~\ref{eq:factorization}, the $v_{\rm n}^{\rm f}$ 
with $1<\ptass<2$\GeVc are extracted as a function of \pttrg\ for various
centrality classes. The associated particle \ptass\ range of 1-2~\GeVc 
is chosen to minimize the non-flow effects. More details of this procedure
can be found in Ref.~\cite{HIN-11-006}.
Various orders of flow harmonics, $v_{\rm n}^{\rm f}$, extracted
from the long-range ($2<|\Delta\eta|<4$) dihadron azimuthal 
correlations for $1<\pttrg<2$~\GeVc
and $1<\ptass<2$~\GeVc, are shown in Fig.~\ref{fig:figure8}
as a function of \npart\ in each centrality bin.
$v_{\rm 2}^{\rm f}$ shows a strong dependence on the collision centrality, 
whereas $v_{\rm 3}^{\rm f}$ is only weakly centrality dependent for \npart\ $> 50$.
This behavior is consistent with the expectation that $v_{\rm 3}^{\rm f}$ is
mainly driven by fluctuations of the initial geometry.
Recent theoretical progress has demonstrated that higher order harmonics are more
sensitive to the finer structure of the medium 
(or finite mean-free-path), and thus the shear viscosity~\cite{Alver:2010dn}.
Therefore, measurements of $v_{\rm n}^{\rm f}$ up to higher orders provide 
important input to better constrain the viscous property of the medium, 
and also shed light on the initial condition of the colliding system.

In summary, the CMS detector at the LHC has been used to measure
angular correlations between charged particles
in $\Delta \eta$ and $\Delta \phi$ up to $|\Delta \eta| \approx 4$
and over the full range of $\Delta \phi$ in pp collisions at \roots\ = 7\TeV
PbPb collisions at \rootsNN\ = 2.76\TeV. A long-range
ridge-like structure is observed at the near-side ($\Delta\phi \approx 0$)
for the first time in pp collisions. Comprehensive studies of the ridge correlation structure in high 
multiplicity pp events, as a function of event multiplicity and particle 
transverse momentum are presented. In PbPb, the extracted
2-D associated yield distributions show a variety of characteristic features
that are not present in minimum bias pp interactions.
Short- and long-range azimuthal correlations have been studied as a function
of the transverse momentum of the trigger particles. The
observed long-range ridge-like structure is most evident in the intermediate transverse momentum
range, $2 < \pttrg < 6$\GeVc, and decreases to almost zero for \pttrg\ above 10--12\GeVc.
A Fourier decomposition of the 1-D $\Delta\phi$-projected
correlation functions in the ridge region ($2 < |\Delta\eta| < 4$) has been presented.
Higher order flow harmonics are extracted as a function of centrality and transverse 
momentum, providing essential insights to the viscosity and initial condition of the PbPb collision system.
The very broad solid-angle coverage of the CMS detector and the 
statistical accuracy of the sample analyzed provide 
significantly improved observations of short- and long-range particle 
correlations over previously available measurements.

%\section*{References}
%\bibliography{Correlations_WeiLi_qm2011_proceeding}

\end{document}